# Potentially Information-theoretic Secure Y00 Quantum Stream Cipher with Limited Key Lengths beyond One-Time Pad

Takehisa Iwakoshi*

Quantum ICT Research Institute, Tamagawa University, 6-1-1, Tamagawa-Gakuen, Machida, Tokyo, Japan

**Abstract**. The previous work showed that the Y00 protocol could stay secure with the eavesdropper's guessing probability on the secret keys being strictly less than one under an unlimitedly long known-plaintext attack with quantum memory. However, an assumption that at least a fast correlation attack is completely disabled by irregular mapping. The present study shows that the Y00 protocol can be information-theoretic secure under any quantum-computational crypto-analyses if the Y00 system is well designed. The Y00 protocol directly encrypts messages with short secret keys expanded into pseudo-random running keys unlike One-Time Pad. However, it may offer information-theoretic security beyond the Shannon limit of cryptography.

**Keywords**: Quantum Cryptography, Information-Theoretic Security, Quantum Optics, Quantum Detection Theory

*Takehisa Iwakoshi, E-mail: t.iwakoshi@lab.tamagawa.ac.jp, TEL.: +81-42-739-8652, FAX: +81-42-739-8663

## 1   Introduction

Since the invention of the first concept of quantum key distribution (QKD), it has been one of the centers of attention in the topic of whether information-theoretic (IT) secure communication is realizable using the laws of quantum physics.

Around year 2000, the Y00 protocol, which was originally called $\alpha\eta$, was proposed by Yuen [1–4] for affinity to existing high-speed and long-distance optical communication infrastructure [7–10]. However, since the fast correlation attack (FCA) on the Y00 protocol was found [11, 12], the Y00 protocol has been believed to be computationally secure, while QKDs are IT secure, even after "irregular mapping" was equipped on Y00 systems as a countermeasure to the FCA [12, 13].

The previous work [14] showed that the eavesdropper "Eve" could not guess the correct secret keys with a probability of one even under an unlimitedly long known-plaintext attack (KPA) with the assistance of quantum memory to utilize the quantum and classical multiple-hypotheses testing theory [15–17]. However, the previous work still assumed that the Y00 system is immune only to the FCA, and no security guarantee against unknown computational attacks exists.

The FCA scheme is generalized herein, and a framework of whether the Y00 protocol can be IT secure against unlimited-resource quantum computers is proposed.

## 2 Brief Description of Security Evaluation of Y00 Protocol in the Previous Work

Although the detailed descriptions are in the previous work [14], this section describes the differences between conventional stream ciphers and quantum-noise-randomized stream ciphers such as the Y00 protocol.

*2.1 Brief Description of Conventional Mathematical Stream Ciphers*

If the KPA is longer than the period of the pseudo-random stream key $s$ generated from the initial key $k$, it completely reveals $s$. Eve then recovers $k$ no matter how complex the key expansion algorithm is because the key expansion is a deterministic process. Hence, she can recover $k$ from $s$ when she knows the algorithm. In terms of conditional probabilities,

$$\Pr(s|c,x) = \Pr(k|c,x) = 1. \tag{1}$$

*2.2 Brief Description of Security of Y00 protocol*

When Eve can KPA longer than the least common multiple (LCM) of the pseudo-random number generators $T_{\mathrm{LCM}}$ equipped in a Y00 system, she will launch an optimal measurement to guess the most probable shared keys. Accordingly, the quantum detection theory for multiple-hypotheses testing is required to evaluate the security of the Y00 protocol in such a case.

To start the Y00 protocol, legitimate users "Alice" and "Bob" must share secret keys $k$ and $\Delta k$. They then expand $k$ and $\Delta k$ into key streams $s$ and $\Delta x$ using the common PRNGs equipped in each transmitter/receiver. Subsequently, *s* is chopped to every $\log_2 M$ bit to form an *M*-ary string *s*(*t*) of time slot *t*, while a message bit *x*(*t*) is encoded into a coherent state $|\alpha[m(t)]\rangle$ as follows using *s*(*t*):

$$m(t) := \text{Map}[s(t)] + M\left(\text{Map}[s(t)] + x(t) + \Delta x(t) \bmod 2\right). \tag{2}$$

Map[*s*(*t*)] is a projection from *s*(*t*) to Map[*s*(*t*)] ∈ {0, 1, 2, 3, … , *M* − 1}. Therefore, the message bit *x*(*t*) ∈ {0, 1} corresponds to a set of quantum states $\{|\alpha[m(t)]\rangle, |\alpha[m(t)+M]\rangle\}$ for even number Map[*s*(*t*)] + Δ*x*(*t*); otherwise, $\{|\alpha[m(t)+M]\rangle, |\alpha[m(t)]\rangle\}$. In contrast, Bob's receiver sets an optimal threshold(s) to discriminate the set of quantum states. Therefore, he decodes *x*(*t*) because he knows Map[*s*(*t*)], thanks to the common PRNG and the shared (***k***, Δ***k***). Meanwhile, eavesdropper Eve has to discriminate the 2*M*-ary signals hidden under the overlapping quantum and classical noise because she does not know whether Map[*s*(*t*)] + Δ*x*(*t*) is even or odd; hence, *x*(*t*) neither.

Eve obtains the coherent states separated from a beam-splitter ρ′[*m*(*t*)] and stores its time sequence in her quantum memory. The quantum sequence ρ′(***x***, ***s***, Δ***x***) with the splitting ratio η is denoted as follows:

$$\rho'(\boldsymbol{x},\boldsymbol{s},\Delta\boldsymbol{x}) := |\eta\alpha(\boldsymbol{x},\boldsymbol{s},\Delta\boldsymbol{x})\rangle\langle\eta\alpha(\boldsymbol{x},\boldsymbol{s},\Delta\boldsymbol{x})| = \bigotimes_{t=0}^{T-1} |\eta\alpha[m(t)]\rangle\langle\eta\alpha[m(t)]|. \tag{3}$$

Note that a set of (***s***, Δ***x***) ∈ (***S***, Δ***X***) is generated from (***k***, Δ***k***) ∈ (***K***, Δ***K***). Therefore, only $2^{|K|+|\Delta K|}$ patterns of single sequences exist, although the number of single levels is 2*M*, and the period of KPA is *T*. Hence, what Eve needs is not a 2*M*·*T*-ary quantum decision theory, but a $2^{|K|+|\Delta K|}$-ary one no matter how long the lengths of ***s*** and Δ***x*** are. The main problem is whether Eve can determine the correct (***s***, Δ***x***) in the LCM of the periods of (***s***, Δ***x***) denoted as $T_{\text{LCM}}$ like in the case of the conventional stream cipher. In the quantum detection theory, the expected probability in successfully obtaining the correct keys is denoted by Γ, while *W* is a Hermitian risk operator. *E* is Eve's optimal measurement operator used to minimize Γ [14–16].

$$W(x, s', \Delta x') := \sum_{(s,\Delta x)\in(S,\Delta X)} \Pr(s, \Delta x)(-\delta_{s,s'}\delta_{\Delta x,\Delta x'})\rho'(x,s,\Delta x)$$
$$= -\Pr(s', \Delta x')|\eta\alpha(x,s',\Delta x')\rangle\langle\eta\alpha(x,s',\Delta x')| \quad (4)$$

$$\Gamma := \sum_{(s,\Delta x)\in(S,\Delta X)} E(s,\Delta x|x)W(x,s,\Delta x) = \sum_{(s,\Delta x)\in(S,\Delta X)} W(x,s,\Delta x)E(s,\Delta x|x). \quad (5)$$

$$\sum_{(s,\Delta x)\in(S,\Delta X)} E(s,\Delta x|x) = I. \quad (6)$$

$$E(s,\Delta x|x) := |(s,\Delta x|x)\rangle\langle(s,\Delta x|x)|. \quad (7)$$

Once a set of the optimal measurement in (6) is determined, from the Cauchy–Schwarz inequality,

$$-\mathrm{tr}\,\Gamma = \sum_{(s,\Delta x)\in(S,\Delta X)} \Pr(s,\Delta x)|\langle\eta\alpha(x,s,\Delta x)|(s,\Delta x|x)\rangle|^2$$
$$\leq \left[\sum_{(s,\Delta x)\in(S,\Delta X)} \Pr(s,\Delta x)^2\right]^{1/2} \left[\sum_{(s,\Delta x)\in(S,\Delta X)} |\langle\eta\alpha(x,s,\Delta x)|(s,\Delta x|x)\rangle|^4\right]^{1/2}$$
$$= \left[\sum_{(s,\Delta x)\in(S,\Delta X)} |\langle\eta\alpha(x,s,\Delta x)|(s,\Delta x|x)\rangle|^4\right]\left[\sum_{(s,\Delta x)\in(S,\Delta X)} |\langle\eta\alpha(x,s,\Delta x)|(s,\Delta x|x)\rangle|^2\right]^{-1} < 1 \quad (8)$$

Therefore, even an unlimitedly long KPA will not reveal the shared key *k* in the Y00 protocol because of the unavoidable quantum noise represented by the term of:

$$\Pr(s',\Delta x'|x,s,\Delta x) = |\langle\eta\alpha(x,s,\Delta x)|(s',\Delta x'|x)\rangle|^2 > 0. \quad (9)$$

## 3 Generalization of Fast Correlation Attack and Corresponding Irregular Mapping

The previous work [14] and the previous section briefly showed that the Y00 protocol can be secured against an unlimitedly long KPA. However, Map[·] is assumed as a well-designed irregular mapping to prevent FCA. Furthermore, whether the Y00 protocol is secure against any mathematical crypto-analyses has not yet been reported. This section revisits the literature [11–13] to determine what conditions are required to show that the Y00 protocol can potentially be IT secure.

*3.1 Generalization of Fast Correlation Attack*

The fundamental idea of the attack in the literature [11, 12] is as follows: Eve observes the key streams ($s'$, $\Delta x'$) that are erroneous compared to the correct ($s$, $\Delta x$) because of the quantum noise. Some of the bits in ($s'$, $\Delta x'$) are not covered by the quantum noise if Map[·] is not well designed. Therefore, she can find the most likely ($s$, $\Delta x$) that corresponds to the initial keys ($k$, $\Delta k$).

In this study, their idea is generalized as follows:

1. The original FCA was launched on a Y00 system that employed Linear Feedback Shift Register (LFSR). No specification on the PRNGs is given herein.
2. Map[·] is not specified; however, some of the bits in $s(t)$ may not be covered by quantum noise.
3. Eve guesses the most likely ($s$, $\Delta x$) by a collective measurement on her quantum memory, different from an individual measurement on each signal in the original FCA.
4. All analyses are done in terms of basic probability and Shannon entropy to show that the following strategy is independent of computational complexity.

In analyzing the situation, consider the measurement operator in (10), where Eve's measurement has an error pattern $e$.

$$E(e\,|\,x) := E((s',\Delta x') - (s,\Delta x) \bmod 2 \,|\, x) = |(e\,|\,x)\rangle\langle(e\,|\,x)| \,. \tag{10}$$

Rewriting (10) in a tensor product in each time slot $t$ similar to (3),

$$\sum_{e \in \{0,1\}^{T_{\text{LCM}}}} E(e\,|\,x) = \bigotimes_{t=1}^{T_{\text{LCM}}} \sum_{e(t) \in \{0,1\}^{|e(t)|}} |(e\,|\,x;t)\rangle\langle(e\,|\,x;t)| = \bigotimes_{t=1}^{T_{\text{LCM}}} I(t) = I \,. \tag{11}$$

$$\Pr\nolimits_{\text{Suc.}}(s,\Delta x\,|\,x,s,\Delta x) = \sum_{e \in \Omega_A} \left|\langle\eta\alpha(x,s,\Delta x)|(e\,|\,x)\rangle\right|^2 \,. \tag{12}$$

$\Omega_A$ is Eve's acceptable range of $e$ to recognize that the measured sequence belongs to $(s, \Delta x)$. Let the length of $e(t)$ be $|e(t)| = |\Delta x(t)| + |s(t)| = 1 + \log_2 M = \log_2 (2M)$ bit. Then,

$$\begin{aligned}\Pr\nolimits_{\text{Suc.}}(s, \Delta x \mid x, s, \Delta x) &= \sum_{e \in \Omega(e)} \prod_{t=1}^{T_{\text{LCM}}} \left| \langle \eta \alpha [m(t)] | (e \mid x; t) \rangle \right|^2 \\ &= \sum_{n(\mathbf{0}) = N(\mathbf{0})}^{N} \sum_{n(e \neq \mathbf{0}) = 0}^{N(e \neq \mathbf{0})} N! \prod_{e \in \{0,1\}^{|e|}} [n(e)!]^{-1} \left| \langle \eta \alpha [m] | (e \mid x; e) \rangle \right|^{2n(e)} . \end{aligned} \quad (13)$$

$$n(\mathbf{0}) \geq N(\mathbf{0}) := N(\mathbf{0}) - \sum_{e' \neq \mathbf{0}} N(e'), \quad (14)$$

where, $N := T_{\text{LCM}} / [\log_2 (2M)]$, and $n(e)$ is the number of $e(t) \neq \mathbf{0}$ appearing in $T_{\text{LCM}}$, and its upper-bound is $N(e)$ such that Eve recognizes that her erroneous measurement result belongs to $(s, \Delta x)$.

The upper-bound of (13) is

$$\begin{aligned}&\log_2 \Pr\nolimits_{\text{Suc.}}(s, \Delta x \mid x, s, \Delta x) \\ &= \log_2 \sum_{n(\mathbf{0}) = N(\mathbf{0})}^{N} \sum_{n(e \neq \mathbf{0}) = 0}^{N(e \neq \mathbf{0})} N! \prod_{e \in \{0,1\}^{|e|}} [n(e)!]^{-1} \Pr(e \mid x, s, \Delta x; e)^{n(e)}, \\ &\leq N \left( H_2 [N(e \neq \mathbf{0})/N] - H_2 [\Pr(e \neq \mathbf{0} \mid x, s, \Delta x; e \neq \mathbf{0})] \right)\end{aligned} \quad (15)$$

where, $H_2[\cdot]$ is Shannon binary entropy, and $e = \mathbf{0}$ indicates that all observed bits are correct after Eve's measurement. The derivations of (14) and (15) are presented in the Appendix.

If we request $\Pr_{\text{Suc.}}(s, \Delta x \mid s, \Delta x, x) = 1$, then the necessary condition is

$$H_2 [N(e \neq \mathbf{0})/N] \geq H_2 [\Pr(e \neq \mathbf{0} \mid x, s, \Delta x; e \neq \mathbf{0})]. \quad (16)$$

The original claim on the FCA [11, 12] is satisfied when for a certain $e \neq \mathbf{0}$,

$$\Pr(e \mid x, s, \Delta x; e) = 1. \quad (17)$$

As (15)–(17) indicate, $N(e)/N = 1$ is allowed. $\Pr_{\text{Suc.}}(s, \Delta x \mid s, \Delta x, x) = 1$ is then obtained. For $e = \mathbf{1}$ in (17), Eve knows that all bits are wrong, which is equivalent to Eve perfectly obtaining the correct bits.

In this way, the original claims in the literature [11, 12] are recovered. Hence, the generalized FCA on the arbitral PRNGs in the Y00 systems is given in terms of information theory under Eve's optimal quantum measurement without any computational assumptions.

*3.2 Requirements in Irregular Mapping against Generalized Fast Correlation Attack*

To implement a secure Y00 system such that $\Pr_{\text{Suc.}}(s, \Delta x \mid s, \Delta x, x) \leq \varepsilon$ with a desirable $\varepsilon$, which was shown in the previous work with some numerical examples [14], for $e \neq 0$,

$$N\left(H_2\left[N(e)/N\right] - H_2\left[\Pr(e \mid x, s, \Delta x; e)\right]\right) \leq \log_2 \varepsilon. \tag{18}$$

Therefore,

$$H_2\left[N(e)/N\right] - N^{-1}\log_2 \varepsilon \leq H_2\left[\Pr(e \mid x, s, \Delta x; e)\right]. \tag{19}$$

Hence, for a corresponding $\delta(\varepsilon^{1/N}, e)$ that satisfies (19),

$$N(e)/N + \delta\left(\varepsilon^{1/N}, e\right) \leq \Pr(e \mid x, s, \Delta x; e). \tag{20}$$

By summing (20) over $e \neq 0$,

$$0 \leq \Pr(0 \mid x, s, \Delta x; 0) \leq N(0)/N - \sum_{e \neq 0} \delta\left(\varepsilon^{1/N}, e\right). \tag{21}$$

The last inequality is what the irregular mapping in the Y00 system requires against the generalized FCA. If such an irregular mapping exists, the Y00 system can be IT secure, similar to the combination of the QKD and the One-Time Pad (OTP).

*3.3 Numerical Examples*

To show a numerical example, suppose that TinyMT [18] are equipped as PRNGs for $(s, \Delta x)$ in the following Y00 system. The seed key length of TinyMT is 127 bits; therefore, $\Pr(k, \Delta k) = \Pr(s, \Delta x) = (2^{127}-1)^{-2}$ for equally possible initial keys, and $T_{\text{LCM}} = 2^{127}-1$ because the period of $\Delta x$ is $2^{127}-1$, while the period of $s$ is $(2^{127}-1)/(\log_2 M)$. Then, $N = (2^{127}-1)/(\log_2 2M)$. Set $\varepsilon = 2^{-64}$ [14] and $N(e)/N = 1/(2M)$ for any $e$ by designing $\{\Pr(e \mid x, s, \Delta x; e)\}$.

The numerical solutions are as follows:

$$\Pr(e \mid x, s, \Delta x; e) \geq 1/(2M) + 3.762 \times 10^{-37} \text{ for } e \neq \mathbf{0}. \tag{22}$$

$$\Pr(\mathbf{0} \mid x, s, \Delta x; \mathbf{0}) \leq 1/(2M) - 1.540 \times 10^{-33}. \tag{23}$$

Note that neither of LFSR nor TinyMT is safe as cryptologic primitives. However, these PRNGs can potentially satisfy IT security when they are equipped in Y00 systems.

However, the optimal quantum measurement operators in (10) must be obtained to perform an exact numerical simulation. It depends on the design of the Y00 system, including its constellation diagram, key-expansion algorithm, and design of irregular mapping. $\{\Pr(e \mid x, s, \Delta x; e)\}$ is then obtained from the above measurement operators. $\{N(e)/N\}$ is given by (25) in the Appendix. Whether the designed Y00 system is IT secure or not is then shown.

*3.4 Principle of IT Secure Y00 Systems*

For a known-plaintext $x$, Eve tries to find a set of shared key-streams ($s$, $\Delta x$), which correspond to the set of shared secret keys ($k$, $\Delta k$). However, if the irregular mapping is well designed, the key streams ($s$, $\Delta x$) are hidden under the quantum noise. Therefore, it is essentially same that the key streams are hidden by OTP with almost IID probability distribution as (22) and (23) because all error patterns would appear almost the same probability.

## 4   Conclusions

This study derived a condition wherein the Y00 protocol would potentially be information-theoretic secure against an eavesdropper who has unlimited computational power with quantum memory, similar to the claim as regards the combination of QKD and One-Time Pad. However, this study has not shown whether such a condition exists. Only one rough estimation was performed in the numerical example section, and Eve's optimal quantum measurement operators and Y00 system design must be derived. Although the numerical example was not precisely given, it showed a principle of how the Y00 protocol would be information-theoretic secure.

## 5  Appendix

(*s*, Δ*x*) and (*s′*, Δ*x′*) are abbreviated herein as *r* and *r′*, respectively.

*5.1 Derivation of* (14)

Consider two possibilities, *r* and *r′*. When Eve decides that the obtained result is *r′* based on an error-pattern *e*, but the true condition is *r*, from the Bayes criterion for $e \neq 0$,

$$\Pr(r')\,_N C_{N(e)} \Pr(e\,|\,x,r';e)^{N(e)} \left[1 - \Pr(e\,|\,x,r';e)\right]^{N-N(e)}$$
$$\geq \Pr(r)\,_N C_{N(e)} \Pr(e\,|\,x,r;e)^{N(e)} \left[1 - \Pr(e\,|\,x,r;e)\right]^{N-N(e)}. \tag{24}$$

For the condition that Eve does not make wrong decisions for any *r* and *r′*,

$$N(e) := \min_{r,r'} \text{Ceil}\left[ N + \frac{N\log_2\left[\Pr(e\,|\,x,r;e)/\Pr(e\,|\,x,r';e)\right] + \log_2\left[\Pr(r)/\Pr(r')\right]}{\log_2\left[1/\Pr(e\,|\,x,r;e) - 1\right] - \log_2\left[1/\Pr(e\,|\,x,r';e) - 1\right]} \right]. \tag{25}$$

Moreover,

$$N(e \neq 0) \geq n(e \neq 0) \text{ with } \sum_{e \in \{0,1\}^{|e|}} n(e) = N. \tag{26}$$

Therefore, define condition *N*(**0**) as (14).

*5.2 Derivation of* (15)

From (26), Eve's success probability in making a correct decision is as follows for $e \neq 0$,

$$\Pr\nolimits_{\text{Suc.}}(s,\Delta x|\,x,s,\Delta x) = \sum_{n(\mathbf{0})=N(\mathbf{0})}^{N}\sum_{n(e\neq\mathbf{0})=0}^{N(e\neq\mathbf{0})} N!\prod_{e\in\{0,1\}^{|e|}}\left[n(e)!\right]^{-1}\Pr(e|\,x,s,\Delta x;e)^{n(e)}$$

$$=\sum_{n(e)=0}^{N(e)}{}_{N}C_{n(e)}\Pr(e|\,x,s,\Delta x;e)^{n(e)}\left[1-\Pr(e|\,x,s,\Delta x;e)\right]^{N-n(e)}.\qquad(27)$$

For a given $\Pr(e\,|\,x,s,\Delta x;e)$,

$$\log_2 \Pr(e|\,x,r;e)^{n(e)}\left[1-\Pr(e|\,x,r;e)\right]^{N-n(e)} \leq -NH_2\left[\Pr(e|\,x,r;e)\right].\qquad(28)$$

Meanwhile, for any $N(e)/N$,

$$\log_2 \sum_{n(e)=0}^{N(e)}{}_{N}C_{n(e)} \leq NH_2\left[N(e)/N\right].\qquad(29)$$

Combining (27)–(29),

$$\log_2 \Pr\nolimits_{\text{Suc.}}(s,\Delta x|\,x,s,\Delta x) \leq N\left(H_2\left[N(e)/N\right]-H_2\left[\Pr(e|\,x,s,\Delta x;e)\right]\right).\qquad(30)$$

*References*

1. H. P. Yuen, "KCQ: A new approach to quantum cryptography I. General principles and key generation," http://arxiv.org/abs/quant-ph/0311061v1, (2003).

2. G. A. Barbosa, E. Corndorf, P. Kumar, and H. P. Yuen, "Secure Communication Using Mesoscopic Coherent States," *Phys. Rev. Lett. Vol,* 90, Issue 22 (2003).

3. C. Liang, G. S. Kanter, E. Corndorf, and P. Kumar, "Quantum noise protected data encryption in a WDM network," *IEEE Photonic Technology Letters,* 17, 1573. (2005).

4. H. P. Yuen, "Key generation: Foundations and a new quantum approach," *IEEE Journal on Selected Topics Quantum Electronics,* 15, 1630. (2009).

5. O. Hirota, K. Kato, M. Shoma, and T. S. Usuda, "Quantum key distribution with unconditional security for all-optical fiber network." in *Quantum Communications and Quantum Imaging* (Vol. 5161, pp. 320- 332). International Society for Optics and Photonics. (2004).

6. O. Hirota, M. Sohma, M. Fuse, and K. Kato, "Quantum stream cipher by the Yuen 2000 protocol: Design and experiment by an intensity-modulation scheme," *Physical Review A,* 72, 022335. (2005).

7. Y. Doi, S. Akutsu, M. Honda, K. Harasawa, O. Hirota, S. Kawanishi, K. Ohhata, and K. Yamashita, "360 km field transmission of 10 Gbit/s stream cipher by quantum noise for optical network," in *Proceedings optical fiber communication conference (OFC),* OWC4. (2010).

8. K. Harasawa, O. Hirota, K. Yaashita, M. Honda, K., Ohhata, S. Akutsu, T. Hosoi, and Y. Doi, "Quantum encryption communication over a 192-km 2.5-Gbit/s line with optical transceivers employing Yuen-2000 protocol based on intensity modulation." *Journal of Lightwave Technology,* 29(3), 323–361. (2011).

9. F. Futami, T. Kurosu, K. Tanizawa, K. Kato, S. Suda, and S. Namiki, "Dynamic Routing of Y00 Quantum Stream Cipher in Field-Deployed Dynamic Optical Path Network," *Optical Fiber Communication Conference* (pp. Tu2G-5). Optical Society of America, (2018).

10. F. Futami, K. Guan, J. Gripp, K. Kato, K. Tanizawa, S. Chandrasekhar, and P. J. Winzer, "Y-00 quantum stream cipher overlay in a coherent 256-Gbit/s polarization multiplexed 16-QAM WDM system." *Optics Express*, 25(26), 33338-33349. (2017).

11. S. Donnet, A. Thangaraj, M. Bloch, J. Cussey, J. M. Merolla, and L. Larger, "Security of Y-00 under heterodyne measurement and fast correlation attack." *Physics Letters A*, 356, 406. (2006).

12. M. J. Mihaljević, "Generic framework for the secure Yuen 2000 quantum-encryption protocol employing the wire-tap channel approach." *Physical Review A*, 75, 052334. (2007).

13. T. Shimizu, O. Hirota, and Y. Nagasako, "Running key mapping in a quantum stream cipher by the Yuen 2000 protocol." *Physical Review A*, 77, 034305. (2008).

14. T. Iwakoshi, "Guessing probability under unlimited known-plaintext attack on secret keys for Y00 quantum stream cipher by quantum multiple hypotheses testing," *Optical Engineering* 57.12 126103, (2018).

15. C. W. Helstrom, "Quantum detection and estimation theory." *Journal of Statistical Physics,* 1(2), 231- 252. (1969).

16. H. P. Yuen, R. Kennedy, and M. Lax, "Optimum testing of multiple hypotheses in quantum

detection theory," *IEEE Transactions on Information Theory*, **21**(2), 125-134, (1975).

17. H. L.Van Trees, K. L. Bell, and Z. Tian, "Detection, estimation, and modulation theory, part I: detection, estimation, and linear modulation theory: 2nd Edition" John Wiley & Sons, (2004).

18. M. Saito and M. Matsumoto, "Tiny Mersenne Twister," http://www.math.sci.hiroshima-u.ac.jp/~m-mat/MT/TINYMT/index.html June, (2011)